\documentclass[aps,prl,twocolumn,showpacs,preprintnumbers,amsmath,superscriptaddress,amssymb,floats,nofootinbib]{revtex4}

\usepackage[usenames]{color}
\usepackage{graphicx}
\begin{document}

\def\Journal#1#2#3#4{{#1} {\bf #2}, #3 (#4)}
\def\NCA{\rm Nuovo Cimento}
\def\NPA{{\rm Nucl. Phys.} A}
\def\NIM{\rm Nucl. Instrum. Methods}
\def\NIMA{{\rm Nucl. Instrum. Methods} A}
\def\NPB{{\rm Nucl. Phys.} B}
\def\PLB{{\rm Phys. Lett.}  B}
\def\PRL{\rm Phys. Rev. Lett.}
\def\PRD{{\rm Phys. Rev.} D}
\def\PRC{{\rm Phys. Rev.} C}
\def\ZPC{{\rm Z. Phys.} C}
\def\JPG{{\rm J. Phys.} G}


\title{Near-threshold Photoproduction of $\phi$ Mesons from Deuterium}

\newcommand*{\DUKE}{Duke University, Durham, North Carolina 27708}
\newcommand*{\DUKEindex}{1}
\affiliation{\DUKE}
\newcommand*{\CALTECH}{Kellogg Radiation Laboratory, California Institute of Technology, California 91125}
\newcommand*{\CALTECHindex}{2}
\affiliation{\CALTECH}
\newcommand*{\OHIOU}{Ohio University, Athens, Ohio  45701}
\newcommand*{\OHIOUindex}{3}
\affiliation{\OHIOU}
\newcommand*{\JLAB}{Thomas Jefferson National Accelerator Facility, Newport News, Virginia 23606}
\newcommand*{\JLABindex}{4}
\affiliation{\JLAB}
\newcommand*{\SCAROLINA}{University of South Carolina, Columbia, South Carolina 29208}
\newcommand*{\SCAROLINAindex}{5}
\affiliation{\SCAROLINA}
\newcommand*{\NOWMIT}{Massachusetts Institute of Technology,
  Cambridge, Massachusetts  02139-4307}
\newcommand*{\NOWMITindex}{6}
\affiliation{\NOWMIT}

\newcommand*{\ANL}{Argonne National Laboratory, Argonne, Illinois 60441}
\newcommand*{\ANLindex}{7}
\affiliation{\ANL}
\newcommand*{\ASU}{Arizona State University, Tempe, Arizona 85287-1504}
\newcommand*{\ASUindex}{8}
\affiliation{\ASU}
\newcommand*{\UCLA}{University of California at Los Angeles, Los Angeles, California  90095-1547}
\newcommand*{\UCLAindex}{9}
\affiliation{\UCLA}
\newcommand*{\CSUDH}{California State University, Dominguez Hills, Carson, CA 90747}
\newcommand*{\CSUDHindex}{10}
\affiliation{\CSUDH}
\newcommand*{\CANISIUS}{Canisius College, Buffalo, NY}
\newcommand*{\CANISIUSindex}{11}
\affiliation{\CANISIUS}
\newcommand*{\CMU}{Carnegie Mellon University, Pittsburgh, Pennsylvania 15213}
\newcommand*{\CMUindex}{12}
\affiliation{\CMU}
\newcommand*{\CUA}{Catholic University of America, Washington, D.C. 20064}
\newcommand*{\CUAindex}{13}
\affiliation{\CUA}
\newcommand*{\SACLAY}{CEA, Centre de Saclay, Irfu/Service de Physique Nucl\'eaire, 91191 Gif-sur-Yvette, France}
\newcommand*{\SACLAYindex}{14}
\affiliation{\SACLAY}
\newcommand*{\CNU}{Christopher Newport University, Newport News, Virginia 23606}
\newcommand*{\CNUindex}{15}
\affiliation{\CNU}
\newcommand*{\UCONN}{University of Connecticut, Storrs, Connecticut 06269}
\newcommand*{\UCONNindex}{16}
\affiliation{\UCONN}
\newcommand*{\EDINBURGH}{Edinburgh University, Edinburgh EH9 3JZ, United Kingdom}
\newcommand*{\EDINBURGHindex}{17}
\affiliation{\EDINBURGH}
\newcommand*{\FU}{Fairfield University, Fairfield CT 06824}
\newcommand*{\FUindex}{18}
\affiliation{\FU}
\newcommand*{\FIU}{Florida International University, Miami, Florida 33199}
\newcommand*{\FIUindex}{19}
\affiliation{\FIU}
\newcommand*{\FSU}{Florida State University, Tallahassee, Florida 32306}
\newcommand*{\FSUindex}{20}
\affiliation{\FSU}
\newcommand*{\GWUI}{The George Washington University, Washington, DC 20052}
\newcommand*{\GWUIindex}{21}
\affiliation{\GWUI}
\newcommand*{\ISU}{Idaho State University, Pocatello, Idaho 83209}
\newcommand*{\ISUindex}{22}
\affiliation{\ISU}
\newcommand*{\INFNFR}{INFN, Laboratori Nazionali di Frascati, 00044 Frascati, Italy}
\newcommand*{\INFNFRindex}{23}
\affiliation{\INFNFR}
\newcommand*{\INFNGE}{INFN, Sezione di Genova, 16146 Genova, Italy}
\newcommand*{\INFNGEindex}{24}
\affiliation{\INFNGE}
\newcommand*{\INFNRO}{INFN, Sezione di Roma Tor Vergata, 00133 Rome, Italy}
\newcommand*{\INFNROindex}{25}
\affiliation{\INFNRO}
\newcommand*{\ORSAY}{Institut de Physique Nucl\'eaire ORSAY, Orsay, France}
\newcommand*{\ORSAYindex}{26}
\affiliation{\ORSAY}
\newcommand*{\ITEP}{Institute of Theoretical and Experimental Physics, Moscow, 117259, Russia}
\newcommand*{\ITEPindex}{27}
\affiliation{\ITEP}
\newcommand*{\JMU}{James Madison University, Harrisonburg, Virginia 22807}
\newcommand*{\JMUindex}{28}
\affiliation{\JMU}
\newcommand*{\KNU}{Kyoungpook National University, Daegu 702-701, Republic of Korea}
\newcommand*{\KNUindex}{29}
\affiliation{\KNU}
\newcommand*{\LPSC}{LPSC, Universite Joseph Fourier, CNRS/IN2P3, INPG, Grenoble, France
}
\newcommand*{\LPSCindex}{30}
\affiliation{\LPSC}
\newcommand*{\UNH}{University of New Hampshire, Durham, New Hampshire 03824-3568}
\newcommand*{\UNHindex}{31}
\affiliation{\UNH}
\newcommand*{\NSU}{Norfolk State University, Norfolk, Virginia 23504}
\newcommand*{\NSUindex}{32}
\affiliation{\NSU}
\newcommand*{\ODU}{Old Dominion University, Norfolk, Virginia 23529}
\newcommand*{\ODUindex}{33}
\affiliation{\ODU}
\newcommand*{\RPI}{Rensselaer Polytechnic Institute, Troy, New York 12180-3590}
\newcommand*{\RPIindex}{34}
\affiliation{\RPI}
\newcommand*{\URICH}{University of Richmond, Richmond, Virginia 23173}
\newcommand*{\URICHindex}{35}
\affiliation{\URICH}
\newcommand*{\ROMAII}{Universita' di Roma Tor Vergata, 00133 Rome Italy}
\newcommand*{\ROMAIIindex}{36}
\affiliation{\ROMAII}
\newcommand*{\MSU}{Skobeltsyn Nuclear Physics Institute, Skobeltsyn Nuclear Physics Institute, 119899 Moscow, Russia}
\newcommand*{\MSUindex}{37}
\affiliation{\MSU}
\newcommand*{\UNIONC}{Union College, Schenectady, NY 12308}
\newcommand*{\UNIONCindex}{38}
\affiliation{\UNIONC}
\newcommand*{\UTFSM}{Universidad T\'{e}cnica Federico Santa Mar\'{i}a, Casilla 110-V Valpara\'{i}so, Chile}
\newcommand*{\UTFSMindex}{39}
\affiliation{\UTFSM}
\newcommand*{\GLASGOW}{University of Glasgow, Glasgow G12 8QQ, United Kingdom}
\newcommand*{\GLASGOWindex}{40}
\affiliation{\GLASGOW}
\newcommand*{\WM}{College of William and Mary, Williamsburg, Virginia 23187-8795}
\newcommand*{\WMindex}{41}
\affiliation{\WM}
\newcommand*{\YEREVAN}{Yerevan Physics Institute, 375036 Yerevan, Armenia}
\newcommand*{\YEREVANindex}{42}
\affiliation{\YEREVAN}

\newcommand*{\NOWCUA}{Catholic University of America, Washington, D.C. 20064}
\newcommand*{\NOWJLAB}{Thomas Jefferson National Accelerator Facility, Newport News, Virginia 23606}
\newcommand*{\NOWLANL}{Los Alamos National Laborotory, New Mexico, NM}
\newcommand*{\NOWGWUII}{The George Washington University2, Washington, DC 20052}
\newcommand*{\NOWWM}{College of William and Mary, Williamsburg, Virginia 23187-8795}

\author {X.~Qian}
\affiliation{\DUKE}
\affiliation{\CALTECH}
\author {W.~Chen}
\affiliation{\DUKE}
\author {H.~Gao}
\affiliation{\DUKE}
\author {K.~Hicks}
\affiliation{\OHIOU}
\author {K.~Kramer}
\affiliation{\DUKE}
\author {J.M.~Laget} 
\affiliation{\JLAB}
\author {T.~Mibe}
\affiliation{\OHIOU}
\author {Y.~Qiang}
\affiliation{\DUKE}
\author {S.~Stepanyan}
\affiliation{\JLAB}
\author {D.J.~Tedeschi}
\affiliation{\SCAROLINA}
% ...

\collaboration{The CLAS Collaboration (2010/12/13)}

\noaffiliation

\date{\today}

\begin{abstract}

We report the first, kinematically-complete measurement of the differential
cross section  of $\phi$-meson photoproduction from deuterium 
near the production 
threshold for a proton using the CLAS detector and a tagged-photon beam in Hall B at Jefferson Lab.
The measurement was carried out by a triple coincidence 
detection of a proton, $K^+$ and $K^-$ near the theoretical production threshold of 1.57 GeV. 
The extracted differential cross sections $\frac{d\sigma}{dt}$ for the initial photon energy range of
1.65-1.75 GeV are 
consistent with predictions based on a quasifree mechanism. 
Our finding is different from recent LEPS results on $\phi$-meson photoproduction from deuterium 
in a similar incident photon energy range, but in a 
different momentum transfer region.

\end{abstract}
\pacs{13.60.Le, 24.85.+p, 25.10.+s, 25.20.-x}
\maketitle

The subject of medium modification of properties of (light) 
vector mesons such as 
their masses and widths has been an important subject in nuclear physics, particularly in the context of partial restoration of chiral symmetry~\cite{chiral_restoration}. Among light vector mesons, the $\phi$ meson is 
considered an excellent candidate for the study of medium modification effects due to its narrow, vacuum width of 4.3 MeV/c$^2$. The first evidence for in-medium modification of the $\phi$ meson at normal nuclear density from 
hadronic reactions was reported by the KEK-PS E325 Collaboration~\cite{kek}, in which a mass shift of 3.4\% and a width increase of a factor of 3.6 for $\phi$ momenta around 1 GeV/c from reaction 
$p + A \rightarrow \rho, \omega, \phi + X \rightarrow e^+e^- + X'$ were deduced.

Studies of $\phi$-meson production from nuclei 
have generated new interest in the last a few years motivated largely 
by a number of new and interesting results reported by different experiments.
A number of experiments reported larger values of 
$\phi$-N total cross section extracted from photoproduction of $\phi$-meson 
from nuclear targets than an
upper limit of $\sigma_{\phi N}\simeq 11$ mb~\cite{sibirtsev} obtained using the
$\phi$-meson photoproduction data on the proton and the vector meson dominance
(VMD) model~\cite{Sakurai}. This upper limit is in agreement with the estimate from the additive
quark model~\cite{Lipkin}. The LEPS collaboration~\cite{leps} reported the $A$-dependence of $\phi$-meson
 photoproduction from Li, C, Al, and Cu nuclear targets, and a larger
value of $\sigma_{\phi N}\simeq 35^{+17}_{-11}$ mb was obtained, corresponding to 
a collisionally broadened width of 35 MeV, much larger than that reported by the KEK-PS E325 Collaboration~\cite{kek}.
The CLAS Collaboration also reported larger values of 
$\sigma_{\phi N}$ from 
a combined analysis~\cite{qian} of 
the coherent~\cite{mibe} and incoherent $\phi$-meson photoproduction from deuterium and a value larger 
than 20 mb is favored for $\sigma_{\phi N}$.
Another recent CLAS experiment~\cite{wood} reported a measured 
value in the range of 16-70 mb for 
 $\sigma_{\phi N}$ from $\gamma A \rightarrow \phi X \rightarrow e^+e^- X$.
More recently, the Anke Collaboration at COSY reported~\cite{anke2010} 
a nuclear dependence 
of $A^{0.56 \pm 0.03}$ from $pA \rightarrow \phi X$ in the $\phi$ meson momentum region of 0.6 - 1.6 GeV/c. The comparison of the data with model calculations suggests an in-medium $\phi$ width about an order of magnitude larger than its vacuum value, which is consistent with a larger value of $\sigma_{\phi N}$~\cite{leps} taking into account the experimental uncertainties 
of both experiments. 

In order to determine the medium modification of the properties of  $\phi$ unambiguously, understanding reaction mechanisms of $\phi$-meson production from nuclear targets is crucial because nuclear targets are necessary in experiments studying the medium effect. The deuteron, a loosely bound and the simplest nuclear system, provides an excellent testing ground to study the reaction mechanisms. Such a study has been reported~\cite{wcchang} recently by the LEPS collaboration on the measurement of the incoherent $\gamma d \rightarrow \phi p n$ photoproduction near threshold at forward angles. The nuclear transparency ratio for $\phi$ has been extracted, which shows a large suppression consistent with the nuclear dependence reported~\cite{leps} by the same collaboration previously using $\phi$-meson photoproduction data from heavier nuclear targets. However, 
 a recent theoretical study~\cite{oset} of $\phi$-meson photoproduction from a deuterium target
shows that the contribution from rescattering at the kinematics of the LEPS experiment is 
small compared with the cross section of $\gamma p \rightarrow \phi p$ in free space. 

Another important aspect of $\phi$-meson production from nuclear targets near threshold is related to experimental investigations of the existence of a $\phi-N$ bound state.
The QCD van der Waals interaction, mediated by multi-gluon exchanges, 
is dominant when two interacting 
color-singlet hadrons have no common valence quarks. 
Bound states of $\eta_{c}$ with $^3$He and 
heavier nuclei were predicted~\cite{Brodsky:1990jd,wasson}. Similarly,
one expects that the attractive QCD van der Waals force also
dominates the $\phi$-N interaction.  A bound state of $\phi$-N  
was found to be possible by Gao, Lee, and Marinov~\cite{gao}. Such a bound state
was also predicted by Huang, Zhang, and Yu using a chiral SU(3) quark model 
and the extended chiral SU(3) quark model by solving the Resonant Group Method 
(RGM) equation~\cite{huang}. 
Subthreshold production kinematics have been proposed~\cite{gao} as advantageous
for the search of $\phi-N$ bound states from nuclear targets.

In this paper, we report on the first kinematically-complete measurement of differential 
cross sections from near-threshold production of $\phi$ mesons
from a deuterium target by a triple coincidence detection 
of $K^+$, $K^-$, and proton.
The measurement was carried out using the CLAS 
detector~\cite{CNIM} at Jefferson Lab. 
The incident photon energy range used in this analysis is 1.65 - 1.75 GeV, which
is above the $\phi$-meson photoproduction threshold ($E_\gamma^{thres.} = 1.57$ GeV) from a free proton target. However,
due to the requirement of a triple-coincidence 
detection and the imperfect acceptance of the detector at forward angles, the reconstructed 
$\phi$ event in this analysis originated mostly from photoproduction on a high-momentum 
proton inside the deuteron, and is below the CLAS production threshold for $\gamma p \rightarrow \phi p \rightarrow K^+ K^- p$. 
Such production therefore images subthreshold production of $\phi$-meson from nuclear targets and as such is important for future experimental search for a $\phi$-N bound state.

High statistics data were collected during the CLAS g10 running period~\cite{g10} 
from a 24-cm-long liquid-deuterium target. 
A tagged-photon beam was used, which was generated by a 
3.8-GeV electron beam incident on a gold radiator with a thickness of 
$10^{-4}$ radiation lengths. The photon flux was measured 
by the Hall B photon-tagging system~\cite{sober}.
Two settings of the CLAS magnetic field were used during the experiment, 
corresponding to a low-field setting 
(with a toroidal magnet current I=2250 A) for better forward-angle 
coverage, and a high-field setting (I=3375 A) for better momentum resolution.
The reaction $d(\gamma,\phi p)n$ was measured 
by detecting kaons from the $\phi$-meson decay ($\phi \rightarrow
K^+ K^-$, branching ratio about 0.5), using 
the same data set as in Refs.~\cite{mibe,qian,weinote}.

\begin{figure} 
\begin{center}
\includegraphics[width=90mm]{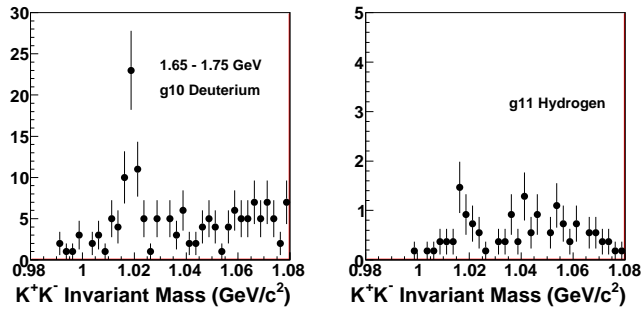}
\caption{The $K^+K^-$ invariant mass distribution from the 
CLAS g10 deuterium data for $E_{\gamma}$ = 1.65-1.75 GeV (left
panel). The $x$-bin size is 2.5 MeV. 
The corresponding distribution from the CLAS g11 hydrogen data set in
the same photon energy range is shown in the right panel.} 
\label{fig:threshold}
\end{center} 
\end{figure}

The $K^+$, $K^-$, and the proton were selected based on the particle charge,
momentum, and time-of-flight   
information. The reaction $d(\gamma,\phi p)n$ was identified in the
missing mass squared distribution by  
applying a $\pm3\sigma$ cut on the missing neutron peak. 
The energy threshold for the $\gamma N \rightarrow \phi N$ reaction is
1.57 GeV.  However, due to the minimum detection threshold for charged particles, 
the CLAS acceptance determined threshold is around 1.75 GeV.
This is demonstrated in our analysis as no $\phi$ events 
at incident photon energies below 1.75 GeV 
can be identified from the g10 hydrogen data set,
\textcolor{black}{which was taken during the g10 running period for
  calibration purposes.} 
This finding is consistent with the hydrogen results from the CLAS g11~\cite{g11} 
high statistics data set. Table~\ref{table:diff} summarizes the differences
between the g10 and g11 experimental settings.
Fig.~\ref{fig:threshold} (left panel) shows the invariant mass distribution 
of the $K^+ K^-$ before \textcolor{black}{the acceptance correction} from
the g10 deuterium data set, where the $\phi$ peak is clearly
visible. Also shown in Fig.~\ref{fig:threshold} (right panel) 
is the corresponding spectrum from the g11 hydrogen data set 
normalized to the g10 integrated luminosity for nucleons.
The yield of the g11 hydrogen is $ \sim 1.5$ events per 2.5 MeV around the
reconstructed $\phi$ meson mass, which is strongly suppressed compared to
that of the g10 deuteron ($\sim 23$ events per 2.5 MeV) due to the energy 
threshold of producing a $\phi$ meson on a nucleon at the kinematic settings 
of g10 and g11.  \textcolor{black}{Therefore, the
    photon energy range used to extract the near-threshold cross section
    for $\phi$-meson photoproduction from deuterium is between 1.65 GeV
    and 1.75 GeV in this work. The chosen photon energy range was
    further confirmed by simulating the $\gamma + p \rightarrow p +
    \phi$ process for the g10 configuration. }  
\begin{table}
  \begin{center}
    \begin{tabular}{|c|ccccc|}
      \hline
      Run &  Magnet & Target &Target&
      Target & Accumulated \\
      Period & Current & Material & Length & Position & Flux\\\hline
      g10 & 2250 (A) & LD$_2$ & 24 cm & -25 cm & 1.266$\times10^{12}$\\
      g10 & 2250 (A) & LH$_2$ & 24 cm & -25 cm & 7.508$\times10^{10}$\\
      g11 &  1930 (A) & LH$_2$ & 40 cm & -10 cm & 4.316$\times10^{12}$\\\hline
    \end{tabular}
    \end{center}
  \caption{A comparison of the g10 and g11 experimental settings. The
  accumulated photon flux is for the $E_{\gamma}$ range of 1.65-1.75 GeV. The ``-10
  cm'' means that the target center of the g11 hydrogen target is 10 cm
  upstream from the nominal center of CLAS.}
\label{table:diff}
\end{table}

Once the reaction $d(\gamma,pK^+K^-)n$ was identified, the number of
$\phi$ mesons was obtained by subtracting the background under the
$\phi$ peak (invariant mass spectrum of the $K^+$ and $K^-$)
in the $\pm 3\sigma$ region (see Fig.~\ref{fig:threshold}). 
The $K^+K^-$ invariant mass distribution 
was fitted using a Breit-Wigner function convoluted with the 
experimental resolution, plus a function to model the background in each
kinematic bin.  The experimental resolution on the missing mass ranged from 
1.2 to 1.7 MeV for different $t$ bins. They were obtained by fitting the 
invariant mass distribution of the Monte-Carlo simulation. 
The background shape was assumed to be~\cite{mibe}:
\begin{eqnarray}
f(x) & = & a \sqrt{x^{2} - (2M_{K})^{2}} + b\left (x^{2}-(2M_{K})^{2}
\right ) ~{\rm for}~ x > 2M_{K} \nonumber\\
f(x) & = & 0~{\rm for}~x<2M_{K},
\end{eqnarray}
where $x$ is the invariant mass of the $K^{+}K^{-}$, $M_K$ is the kaon mass, 
and $a$ and $b$ are fitting parameters. Such a fit was performed separately for 
each $t$ bin, and the $t$ dependence of the background was effectively included in
the fitting parameters $a$ and $b$. In addition, the background was also 
fitted to a straight line. The results from
  fitting these two shapes were compared in order to estimate the systematic
  uncertainties due to the subtraction of the background.

 GEANT3 Monte-Carlo (MC) simulations were carried out to model 
detector efficiencies and resolutions for this reaction channel. 
A quasifree event generator was used for the near-threshold kinematics.
It generated $pK^{+}K^{-}$ three-body events with a random photon 
energy based on the photon energy distribution of the data in the energy 
range of interest. The initial momentum of the nucleon inside the deuteron was
chosen using the Bonn potential wavefunction~\cite{bonn}. 
The spectator nucleon was assumed to be on-shell, whereas the struck nucleon
was assumed to be off-shell before absorbing a photon.
Isospin symmetry was assumed for the $\phi$-meson
photoproduction from the nucleon. The events were then checked to
ensure energy conservation. 
The MC events were generated based on a Breit-Wigner shape of the
resonance centered at the $\phi$ mass of 1.019 GeV$^2$ with a full width 
at half maximum 
(FWHM) of $\Gamma$=4.26 MeV.  
The $\phi$ meson decay angular distribution and cross section are
based on the g11 hydrogen data~\cite{g11dxs}. 

\begin{figure} 
\begin{center}
\includegraphics[width=90mm]{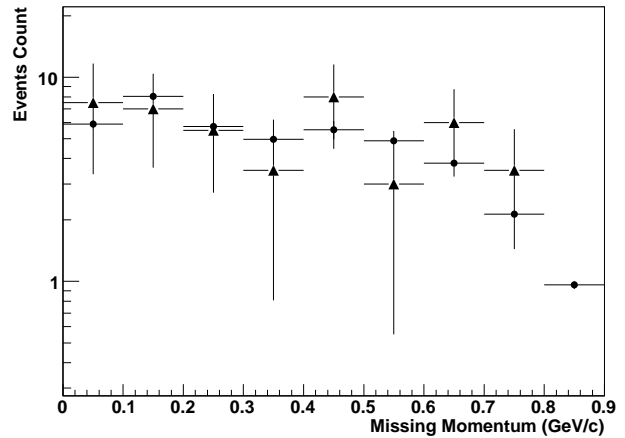}
\caption{Missing momentum distributions are shown for
  data (solid triangles) and MC (solid circles), 
where the MC results are normalized to
  the luminosity of the data. Both $\phi-N$ and $N-N$ FSI are included in
  the Monte Carlo. }
\label{fig:missp}
\end{center} 
\end{figure}

The $\phi$-meson differential photoproduction cross section on a
hydrogen target ($\frac{d\sigma}{dt}$ vs. $t-t_0$) was obtained using 
a fit to the g11 data in $E_{\gamma} = 1.625 -3.775$ GeV range. Here $t$
is the four-momentum transfer squared, $(P_{\phi}-P_{\gamma})^2$, and $t_0$
 is the maximum $t$ value for a given photon energy because $t$ is negative. 
In addition, the
  event generator included the $N-N$ and $\phi-N$ final-state
  interactions (FSIs).  The Jost function  
  approach~\cite{Jost} was used for the $N-N$ FSI. 
 The $\phi-N$ FSI, which was assumed to be incoherent from the original
$\phi$-meson photoproduction process, was modeled based on the
vector meson dominance
(VMD) model, in which the $t$-dependence of the $\phi-N$ elastic
scattering cross section is the same as that of
$\phi$-meson photoproduction. A fitting procedure, in which 
the strength of $N-N$ and $\phi-N$ FSIs were obtained, 
was then used to optimize the $\phi$-meson photoproduction model 
so that the resulting Monte-Carlo distributions match those of the
data. Fig.~\ref{fig:missp} shows the comparison in the missing
momentum distribution for the data (solid triangles) and the MC
(solid circles).   

MC-generated events were used as input to the GEANT3-based CLAS
simulation~\cite{gsim}. They were then reconstructed using the same 
 algorithm as was used for the data.
The acceptance was obtained by the ratio of the number of events 
that passed the analysis cuts to the number of 
generated $\phi$ events. The average differential cross sections were
extracted by dividing the normalized yield by the acceptance. The
differential cross sections were then bin-centered at fixed $t$ values 
with a finite binning correction.

Several sources contribute to the overall systematic uncertainty in
the differential cross section.
The systematic uncertainties associated with particle identification 
and the missing mass cut were  4.2 - 12.9\% and 1.4 -
10.9\%, respectively. These values were determined by varying the corresponding cuts by $\pm 10$\% in each $t$ bin. 
The angular distributions of the $\phi$-meson's decay products 
in its rest frame and the $\rm {cos}(\theta_{\rm {c.m.}})$ distribution 
of the $\phi$ meson were uncertain to within 10\% and 5\%~\cite{g11dxs,prephi}, 
respectively,
leading to 5.2-13.2\% systematic uncertainties. The background obtained
from the non-linear background shape was on average 8\% smaller than
that from the linear background. A conservative 8\% systematic uncertainty is
assigned for the background subtraction procedure. 
The systematic uncertainties due to the effect of FSIs 
were obtained by varying the fitted 
strengths of the $N-N$ FSI and the $\phi-N$ FSI by 30\% and 50\%, 
respectively. The systematic errors vary from 4\% to 17\% for different $t$ bins.
The uncertainty in the photon flux was
5\%~\cite{gflux,weinote}. \textcolor{black}{The uncertainty of the bin-centering
correction was assumed to be 30\% of the size of the correction
based on knowledge of the CLAS acceptance and 
the input cross section model, which is obtained from the g11 
data~\cite{g11dxs}. The absolute size of
bin-centering corrections was between 1\% and 10\%.}
Combined in quadrature, the overall systematic 
errors vary from $18\% - 32\%$ depending on the kinematics.

\begin{figure} 
\begin{center}
\includegraphics[width=90mm]{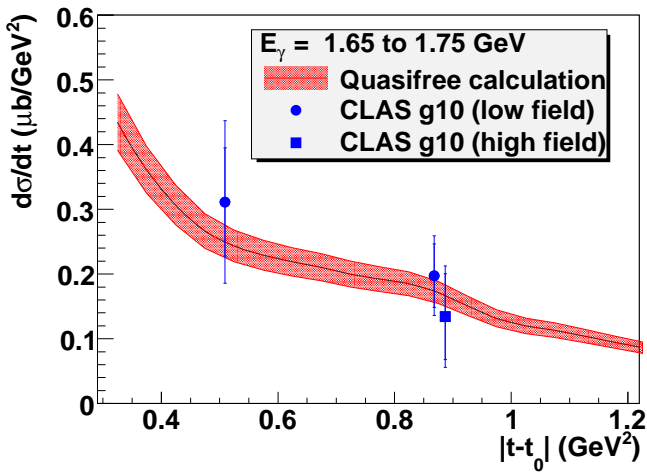}
\caption{(Color online) $\phi$ photoproduction
  differential cross sections from deuterium are plotted as a function
  of $|t-t_0|$. The inner error bars are the statistical
  uncertainties, and the outer error bars are the quadrature sum of
  systematic and statistical uncertainties. } 
\label{fig:subd}
\end{center} 
\end{figure}

\begin{table}
  \begin{center}
    \begin{tabular}{|ccccc|}
      \hline
      $E_{\gamma}$ & $|t-t_{0}|$ & $\frac{d\sigma}{dt}$ & stat. &
      sys. \\
      (GeV) & (GeV$^2$) & ($\mu$b/GeV$^2$) & uncer. & uncer. \\\hline
      1.65-1.75 & 0.509 & 0.31   &     0.084 &      0.094 \\
      1.65-1.75 & 0.887 & 0.20   &     0.049  &     0.037 \\
      1.65-1.75 & 0.924 & 0.13   &     0.066 &      0.041 \\\hline
    \end{tabular}
    \end{center}
  \caption{Tabulated results on the $\phi$-meson photoproduction 
from deuterium for an $E_{\gamma}$ range of  1.65 to 1.75 GeV.}
\label{table:subd}
\end{table}

The $\phi$-meson photoproduction cross sections for 
the deuteron are tabulated in Table~\ref{table:subd} and are plotted as a
function of $|t-t_0|$ in Fig.~\ref{fig:subd} for a photon energy  
range of 1.65-1.75 GeV. The solid circles are results
obtained from the CLAS g10 low-field setting, whereas the solid
square is the result from the high-field setting. \textcolor{black}{The quasifree
calculation is also plotted for comparison together with its
uncertainty (shown as a band in  Fig.~\ref{fig:subd}). This simple
calculation is based on a quasifree picture with the $\phi$-meson
differential  photoproduction cross section from the proton which is based on
the g11 data~\cite{g11dxs}. The principle of this calculation is the
same as that for the event generator used in the MC. The systematic uncertainty for
this calculation is about 10\% due to the uncertainty in the input
cross section. The extracted
differential cross section for $\phi$-meson photoproduction is consistent
with the quasifree calculation within uncertainties. } 
Our finding is consistent with theoretical calculations presented in Ref.~\cite{oset}. However, 
they are not consistent with the recent LEPS results~\cite{wcchang} from deuterim in a similar photon energy range, but 
in a smaller momentum transfer region. The corresponding $|t-t_0|$ range for the LEPS data in a photon energy range of 1.65 to 1.75 GeV 
is less than 0.2 GeV$^2$.  
Future studies both in experiment and in theory are important to clarify 
the situation.

In summary, we have extracted for the first time the differential
cross section  
on $\phi$-meson photoproduction from deuterium from a kinematically-complete measurement below the production 
threshold for the proton accessible on CLAS. The chosen 
incident photon energy range is 1.65-1.75 GeV, which is near the 
1.57 GeV production threshold for protons.  
Our extracted differential cross sections are in agreement 
with predictions from a simple quasifree picture. 
This finding is consistent with the recent theoretical 
study of this reaction~\cite{oset}, though inconsistent with the recent LEPS results in a somewhat 
different kinematic region. Further, our data provide information on $\phi$-meson subthreshold photoproduction cross section from a deuterium target.
Although heavier
nuclear targets will be ideal for future dedicated searches for 
a $\phi$-N bound state, the extracted cross sections from deuterium reported 
in this paper will help provide information on the expected 
production rate of the $\phi$-N bound state from heavier nuclear targets.

We thank helpful comments from W.C. Chang and E. Oset. 
We acknowledge the outstanding efforts of the staff  
of the Accelerator and Physics 
Divisions at Jefferson Lab who made this experiment possible. 
This work was supported in part by the U.S.~Department of Energy, 
the National Science Foundation, 
the Italian Istituto Nazionale di Fisica Nucleare, the French 
Centre National de la Recherche Scientifique and Commissariat \`a 
l'Energie Atomique, and the National Research Foundation of Korea. 
Jefferson Science Associates (JSA) operates the 
Thomas Jefferson National Accelerator Facility for the 
U.S.~Department of Energy under contract DE-AC05-84ER40150.

\end{document}